# Nonlinear Modal Decoupling Based Power System Transient Stability Analysis

Bin Wang, *Member, IEEE*, Kai Sun, *Senior Member, IEEE*, and Xin Xu, *Student Member, IEEE*

*Abstract*—Nonlinear modal decoupling (NMD) was recently proposed to nonlinearly transform a multi-oscillator system into a number of decoupled oscillators which together behave the same as the original system in an extended neighborhood of the equilibrium. Each oscillator has just one degree of freedom and hence can easily be analyzed to infer the stability of the original system associated with one electromechanical mode. As the first attempt of applying the NMD methodology to realistic power system models, this paper proposes an NMD-based transient stability analysis approach. For a multi-machine power system, the approach first derives decoupled nonlinear oscillators by a coordinates transformation, and then applies Lyapunov stability analysis to oscillators to assess the stability of the original system. Nonlinear modal interaction is also considered. The approach can be efficiently applied to a large-scale power grid by conducting NMD regarding only selected modes. Case studies on a 3-machine 9-bus system and an NPCC 48-machine 140-bus system show the potentials of the approach in transient stability analysis for multi-machine systems.

*Index Terms*—Nonlinear modal decoupling, transient stability analysis, energy function, first-integral, Zubov's method.

## I. Introduction

MODERN power systems are sometime operated close to their stability limits. A major function of dynamic security assessment is to check transient stability and dynamic performance of a power system under a severe contingency. Numerical time-domain simulation on "what-if" scenarios is a widely adopted approach by power industry for identification of potential transient instabilities, whose results are accurate but contingency dependent. Compared to numerical approaches, analytical methods for transient stability analysis (TSA), which have been studied for decades like the so-called direct methods based on the Lyapunov theory [1][2], not only judges the stability for a contingency scenario but also provide in-depth understanding of system dynamical behaviors and stability margin information, which is important for ranking contingencies and designing remedial control actions [3].

The main challenges faced by direct methods are the complexity in nonlinear dynamics with a multi-machine power system and the non-existence of an ideal Lyapunov function for a general power system [4]. In addition, detailed power system models are difficult to be considered in direct methods.

Hybrid methods combining the advantages of both numerical and analytical methods have been explored, e.g., by Ref. [5], which applies a direct method to calculate a stability margin index while simulating a power system, and uses early-termination criteria based on this index to save CPU time. Some other hybrid methods, such as the Extended Equal Area Criterion (EEAC) method [6], the SIngle-Machine Equivalent (SIME) method [7], reduce a complex multi-machine system model or its simulated trajectories to a two-machine equivalent system and consequently an equivalent single-machine-infinite-bus (SMIB) system so that direct methods such as the classical equal area criterion can be applied. However, these methods implicitly assume that the instability mode of a multi-machine system can be characterized by an equivalent SMIB system, but which is not always true [8]. For example, it was found that the nonlinear behaviors of a multi-machine system associated with an electromechanical mode before instability happens can be more accurately characterized by a $2^{nd}$ order non-SMIB nonlinear system than by any SMIB system [8].

For a nonlinear, multi-oscillator system, Ref. [8] proposes a methodology, named nonlinear modal decoupling (NMD), to systematically construct as many decoupled nonlinear oscillators as the system's oscillation modes such that each individual oscillator can be analyzed separately to infer the original system's dynamics and stability associated with one mode. By means of a series of coordinates transformations on the original system's state space and mathematical model, the set of decoupled oscillators can be obtained and they together provide a fairly accurate representation of the original system's behaviors subject to both small and large disturbances within an extended region about its equilibrium in the state space.

This paper is the first attempt applying NMD to TSA for practical power system models. For a multi-machine power system, the proposed NMD-based approach for TSA first derives decoupled second-order nonlinear systems, which do not necessarily have to be in a similar form to that of SMIB systems. Then, the approach applies Lyapunov stability analysis to each decoupled system to assess the transient stability of the original multi-machine system. For a large-scale power grid, the proposed approach performs NMD regarding only selected dominant modes to ensure its time performance. The rest of the paper is organized as follows: Section II briefly introduces the NMD method for power systems. Section III presents three methods for estimating the

This work was supported in part by NSF CAREER Award (ECCS-1553863) and in part by the ERC Program of the NSF and DOE under Grant EEC-1041877.

B. Wang, K. Sun and X. Xu are with the University of Tennessee, Knoxville, TN 37996 USA (e-mail:bwang13@vols.utk.edu, kaisun@utk.edu, xxu30@vols.utk.edu).

stability boundary of each mode-decoupled system given by NMD. The application of NMD on power system TSA is introduced in Section IV with the consideration of large power systems. Case studies on both small systems and a large NPCC 48-machine 140-bus system will be presented in Section V to validate the proposed approach. Section VI draws the conclusions and envision the future works.

## II. Nonlinear Modal Decoupling of Power Systems

Given a multi-oscillator system described by $N$ ordinary differential equations, which may represent a power system:

$$\dot{\mathbf{x}} = \mathbf{f}(\mathbf{x}) \qquad (1)$$

where $N$ is an even number, $\mathbf{x}=[x_1, x_2, \ldots, x_N]^T$ is the state vector with the equilibrium at the origin and $\mathbf{f}(\mathbf{x})=[f_1(\mathbf{x}), f_2(\mathbf{x}), \ldots, f_N(\mathbf{x})]^T$ is a smooth vector field. Assume that the Jacobian matrix of $\mathbf{f}(\mathbf{x})$, denoted by $A$, has $N/2$ conjugate pairs of complex eigenvalues $\lambda_1, \lambda_2, \ldots, \lambda_N$. Without loss of generosity, let $\lambda_{2i-1}$ and $\lambda_{2i}$ be a conjugate pair defining the mode $i$.

It has been shown in [8] that, if the resonance does not happen, the system in (1) can always be transformed to a dynamical system as shown in (2) first by a linear transformation that diagonalizes $A$ using its modal matrix (consisting of all right eigenvectors) and then by a series of homogeneous polynomial transformations defined in (3) and (4) with $p=1, 2, \ldots$.

$$\dot{\mathbf{z}}^{(k)} = \mathbf{\Lambda} \cdot \mathbf{z}^{(k)} + \sum_{j=2}^{k} \mathbf{D}_j(\mathbf{z}^{(k)}) + \sum_{j=k+1}^{\infty} \left( \mathbf{D}_j(\mathbf{z}^{(k)}) + \mathbf{C}_j(\mathbf{z}^{(k)}) \right) \qquad (2)$$

$$\mathbf{z}^{(p)} = \mathbf{H}_{p+1}(\mathbf{z}^{(p+1)}) \stackrel{\text{def}}{=} \mathbf{z}^{(p+1)} + \begin{bmatrix} \vdots \\ h_{p+1,2i-1}(\mathbf{z}^{(p+1)}) \\ h_{p+1,2i}(\mathbf{z}^{(p+1)}) \\ \vdots \end{bmatrix} \qquad (3)$$

$$\begin{cases} h_{p+1,2i-1}(\mathbf{z}^{(p+1)}) = \sum_{\alpha=1}^{N} \cdots \sum_{\gamma=\eta}^{N} h_{p+1,2i-1,\alpha\cdots\eta\gamma} \underbrace{z_\alpha^{(p+1)} \cdots z_\gamma^{(p+1)}}_{p+1 \text{ terms in total}} \\ h_{p+1,2i}(\mathbf{z}^{(p+1)}) = \overline{h}_{p+1,2i-1}(\mathbf{z}^{(p+1)}) \end{cases} \qquad (4)$$

If only $k$ terms of (2) are summated, a truncated approximate of (1) is produced. In (2), $\mathbf{z}^{(k)}=[z_1^{(k)}, z_2^{(k)}, \ldots, z_N^{(k)}]^T$ is the state vector of the transformed system having $z_{2i-1}^{(k)}$ and $z_{2i}^{(k)}$ associated with mode $i$, and $\mathbf{D}_j$ and $\mathbf{C}_j$ are vector fields whose elements are weighted sums of the terms of degree $j$ about $\mathbf{z}^{(k)}$. $\mathbf{D}_j$ only contains intra-modal terms, i.e. monomials only about the state variables associated with mode $i$, while $\mathbf{C}_j$ only has inter-modal terms, i.e. monomials each about state variables with different modes.

Note that $h$-coefficients in (4) are not unique. Ref. [8] provides two ways to determine a set of $h$-coefficients: one assumes that each mode-decoupled system is an equivalent SMIB system, which is called "SMIB assumption"; the other one determines $h$-coefficients under a so-called "smaller transfer assumption", which tries to avoid propagating nonlinearities from low order terms to high order terms. As reported in [8], the "small transfer assumption" generates a set of $h$-coefficients leading to a more accurate approximate to the original system than the "SMIB assumption", and hence is adopted throughout this paper.

In (2), all modes are decoupled in the terms of up to order $k$. If we ignore nonlinear terms of orders $\geq k+1$, where modal coupling still remains, a truncated but mode-decoupled system (5) is resulted, where the pair of equations about each individual mode are completely decoupled from the other ($2N-2$) equations about other modes.

$$\dot{\mathbf{z}}_{\text{jet}}^{(k)} = \mathbf{\Lambda} \cdot \mathbf{z}_{\text{jet}}^{(k)} + \sum_{j=2}^{k} \mathbf{D}_j(\mathbf{z}_{\text{jet}}^{(k)}) \qquad (5)$$

Thus, each mode $i$ can be analyzed using only the associated pair of equations, which are rewritten as (6).

$$\begin{cases} \dot{z}_{\text{jet},2i-1}^{(k)} = \lambda_{2i-1} z_{\text{jet},2i-1}^{(k)} + \sum_{\alpha=1}^{N} \sum_{\beta=\alpha}^{N} \mu_{2i-1,\alpha\beta} z_{\text{jet},\alpha}^{(k)} z_{\text{jet},\beta}^{(k)} + \cdots + \\ \qquad \sum_{\alpha=1}^{N} \cdots \sum_{\rho=\gamma}^{N} \mu_{2i-1,\alpha\cdots\rho} \underbrace{z_{\text{jet},\alpha}^{(k)} \cdots z_{\text{jet},\rho}^{(k)}}_{k \text{ terms in total}} \stackrel{\text{def}}{=} f_{2i-1}^{(k)}(\mathbf{z}_{\text{jet}}^{(k)}) \\ \dot{z}_{\text{jet},2i}^{(k)} = f_{2i}^{(k)}(\mathbf{z}_{\text{jet}}^{(k)}) = \overline{f}_{2i-1}^{(k)}(\mathbf{z}_{\text{jet}}^{(k)}) \end{cases} \qquad (6)$$

Note that the two state variables and all coefficients in (6) are complex-valued. It would be more convenient to analyze its equivalent real-valued system, whose two state variables have physical meanings of, e.g., displacement and velocity. For this purpose, the linear transformation (7) is used to transform (6) into (8) [8].

$$\begin{bmatrix} z_{\text{jet},2i-1}^{(k)} \\ z_{\text{jet},2i}^{(k)} \end{bmatrix} = \begin{bmatrix} \lambda_{2i-1} & \lambda_{2i} \\ 1 & 1 \end{bmatrix}^{-1} \begin{bmatrix} w_{2i-1} \\ w_{2i} \end{bmatrix} \qquad (7)$$

$$\begin{cases} \dot{w}_{2i-1} = \upsilon_{i10} w_{2i-1} + \sum_{l=1}^{k} \upsilon_{i0l} w_{2i}^l + \sum_{\substack{j \geq 1, l \geq 0 \\ (j,l) \neq (1,0)}}^{j+l \leq k} \upsilon_{ijl} w_{2i-1}^j w_{2i}^l \\ \dot{w}_{2i} = w_{2i-1} + \sum_{\substack{j \geq 0, l \geq 0}}^{2 \leq j+l \leq k} v_{ijl} w_{2i-1}^j w_{2i}^l \end{cases} \qquad (8)$$

**Remarks:** Errors are introduced in two of steps from (1) to (8): the first is the Taylor expansion of (1) that introduces truncation errors, and the second is the ignoring of terms of orders $>k$ in (2). Reference **Error! Reference source not found.** shown that the third- or fourth-order truncated Taylor expansion of power system swing equations most likely give conservative stability analysis results, so this paper adopts third-order Taylor expansion for all case studies. Dynamics governed by high-order terms may be ignored in a neighborhood of the stable equilibrium point (SEP); however, when the system state is moving toward its stability boundary, the system (5) or its equivalent (8), will have increased errors.

Ref. [8] also reported that by using (8), analysis on the stability boundary for system (1) regarding the mode of interest becomes much simpler and can be fairly accurate. This enables a new NMD-based analytical approach for TSA. Instead of analyzing the original system (1), the new approach uses a pair of differential equations from the decoupled real-valued system (8) to estimate the stability boundary observed from the two-dimensional state subspace about one mode as well as to estimate dynamical behaviors of system (1) under disturbances by means of the NMD transformations in (2) and (7). The rest of this paper will focus on the NMD-based TSA approach.

## III. Stability Analysis of Mode-Decoupled Systems

This section introduces the identification of the stability boundary of the real-valued mode-decoupled system (8). A time simulation aided stability boundary search algorithm is introduced to provide a reference result. Then, two analytical approaches based on the Lyapunov stability theory are presented [2]. With a few simplifying assumptions, the first approach, called the first integral method, constructs a unique Lyapunov function and the critical energy is calculated at the unstable equilibrium point (UEP) close to the origin. The second approach, called the Zubov's method, does not require any assumptions in the original definition of the Lyapunov function. However, the closed-form solution is not always achievable. Thus, an approximation by a truncated power series solution is adopted along with a clear definition of critical energy. It should be noted that other methods, e.g. Popov's method, can also be adopted, which will be investigated in future.

### A. Time simulation aided stability boundary search

Consider the dynamical system (8). The stability boundary of its SEP is the boundary of the basin of attraction (BOA), which can be estimated based on the following observation: if the system trajectory starting from an initial state converges to the SEP, that initial state is inside the BOA; otherwise, it is outside. The stability boundary can be numerically estimated by the searching algorithm below, similar to the one in [12].

***Step 1:*** In the $w_{2i-1}$-$w_{2i}$ plane, set up $M$ unit vectors with angles (in degree) respectively taking 0, 360/$M$, …, 360 ($M$-1)/$M$, say $n_1$, …, $n_M$. $M = 180$ is used in all studied cases in this paper.
***Step 2:*** Set $s$ to be a small step, e.g. 0.1 used in this paper.
***Step 3:*** Along the direction determined by each of the unit vectors, say $n_j$, use $w_0 = sn_j$ as the initial state to numerically solve (8) over a period of time (5 seconds in this paper) for $w_{2i-1}(t)$ and $w_{2i}(t)$.
***Step 4:*** If $s < \varepsilon$, stop the search and $w_0$ is an estimate of the boundary in direction $n_j$; otherwise, go to step 5. $\varepsilon$ in this stopping criterion defines the error tolerance and takes 0.01 in this paper.
***Step 5:*** If the gap between the maximum and minimum of $w_{2i}(t)$ is >750°, let $w_0 = w_0 - sn_j$ and $s = s/2$; otherwise, go to step 6.
***Step 6:*** let $w_0 = w_0 + sn_j$ and go to step 3.

This numerical searching algorithm can give a fairly accurate estimate when step $s$ and tolerance $\varepsilon$ take very small values. The entire time cost is moderate since the system (8) only has two state variables. With this estimated boundary as a reference, the following will present two analytical approaches for estimating the stability boundary.

### B. First integral

Eq. (9) gives a necessary and sufficient condition for (1) to have a first-integral based Lyapunov function. Unfortunately, there are no general methods for constructing a first-integral based Lyapunov function for nonlinear dynamical systems [2].

$$\sum_{j=1}^{N} \frac{\partial f_j}{\partial x_j} = 0 \qquad (9)$$

With the help of NMD, if the assumptions in (10) and (11) hold, system (8) is transformed to (12), whose first-integral based Lyapunov function can be constructed as in (13).

$$\begin{cases} \upsilon_{ijl} = 0 & \text{for all } j \geq 1, l \geq 1, j+l \leq k, (j,l) \neq (1,0) \\ \nu_{ijl} = 0 & \text{for all } j \geq 0, l \geq 0, 2 \leq j+l \leq k \end{cases} \qquad (10)$$

$$\upsilon_{i10} = 0 \qquad (11)$$

$$\begin{cases} \dot{w}_{2i-1} = \sum_{l=1}^{k} \upsilon_{i0l} w_{2i}^{l} \\ \dot{w}_{2i} = w_{2i-1} \end{cases} \qquad (12)$$

$$\frac{dw_{2i}}{dw_{2i-1}} = \frac{\dot{w}_{2i}}{\dot{w}_{2i-1}} = \frac{w_{2i-1}}{\sum_{j=1}^{k} \upsilon_{ij} w_{2i}^{j}} \Rightarrow V(w_{2i-1}, w_{2i})$$

$$= \frac{w_{2i-1}^{2}}{2} - \int_{0}^{w_{2i}} \sum_{j=1}^{k} \upsilon_{ij} s^{j} ds = \frac{w_{2i-1}^{2}}{2} - \sum_{j=1}^{k} \frac{\upsilon_{ij}}{j+1} w_{2i}^{j+1} \qquad (13)$$

Note that the assumption in (10) changes the nonlinear characteristics of the system, while the assumption (11) forces the oscillation damping to zero, which does not have a significant influence on the stability analysis result. Validated by the numerical studies presented later in subsections V-B and V-C, the coefficients ignored by (10) are always found to be small and ignoring positive damping can keep the stability analysis based on (12) to be conservative.

For each decoupled system in (12), the closest UEPs denoted by $w_{2i,\text{UEP}}$ can be obtained by letting the right hand side be zero and solving the resulting algebraic equations for the roots with the smallest magnitude. Note that there may be one or two (a positive one and a negative one) closest UEPs depending on the order $k$. Then, the critical energy is defined as $V(0, w_{2i}^{*})$, where $w_{2i}^{*}$ is selected as the UEP having a smaller energy. When the systems in (12) has an initial state $(w_{2i-1}(0), w_{2i}(0))$, it is stable if and only if $V(0, w_{2i}^{*}) \geq V(w_{2i-1}(0), w_{2i}(0))$. Therefore, the stability boundary of (8) can be approximated by an equipotential line of (12) with the potential of $V(0, w_{2i}^{*})$, i.e.:

$$V(w_{2i-1}, w_{2i}) = V(0, w_{2i}^{*}) \qquad (14)$$

### C. Zubov's method

A Lyapunov function $V(\mathbf{x})$ for determining the exact stability boundary of the ordinary differential equations in (1) can be constructed by solving the partial differential equation in (15), called the Zubov's equation [13][14]. Note that the mode-decoupled system in (8) is a special case of the general system in (1). Thus, this subsection only applies the Zubov's method to (1). All conclusions drawn are automatically applicable to (8).

$$\sum_{j=1}^{N} \frac{\partial V(\mathbf{x})}{\partial x_j} f_j(\mathbf{x}) = -\varphi(\mathbf{x}) \cdot (1 - V(\mathbf{x})) \qquad (15)$$

where $\varphi(\mathbf{x})$ is a positive definite or semidefinite function of $\mathbf{x}$. Note that $\varphi(\mathbf{x})$ has to be chosen before solving the above equation, and its selection will not influence the resulting stability boundary.

Several theorems (see proofs in [13]) and necessary definitions are briefly reviewed below due to their importance for understanding the Zubov's method based stability analysis.
**Theorem III-C1.** The function $V(\mathbf{x})$ solved from (15) is a Lyapunov function establishing the asymptotic stability of the unperturbed motion at the SEP of the system (1).

">3

**Definition III-C1.** Let $\Omega$ be the set of the initial values $\mathbf{x}_0$ which make up the domain of asymptotic stability of the unperturbed motion at the SEP of the system in (1). Thus, $\Omega$ is the BOA of the system (1).

**Theorem III-C2.** If $\mathbf{x}$ is in $\Omega$, then
$$0 \leq V(\mathbf{x}) < 1 \tag{16}$$

**Theorem III-C3.** The stability boundary of the system in (1), i.e. the boundary of $\Omega$, is the surface defined by $V(\mathbf{x}) = 1$.

Eq. (15) has a closed-form solution only for some special cases, but its solution can be represented in a power series form (17), where $V_j(\mathbf{x})$ represents all homogeneous terms of order $j$ in $\mathbf{x}$. Truncating the terms having orders high than $L$, an approximate of $V(\mathbf{x})$ is given in (18).

$$V(\mathbf{x}) = \sum_{j=2}^{\infty} V_j(\mathbf{x}) \tag{17}$$

$$V(\mathbf{x}) \approx V^{(L)}(\mathbf{x}) = \sum_{j=2}^{L} V_j(\mathbf{x}) \tag{18}$$

**Definition III-C2.** Given $V^{(L)}(\mathbf{x})$ in (18), define the set $\Phi$ and the constant scalar $v^{(L)}$ respectively by (19) and (20).

$$\Phi = \{\mathbf{x} \mid \dot{V}^{(L)}(\mathbf{x}) = 0\} \tag{19}$$

$$v^{(L)} = \min\{V^{(L)}(\mathbf{x}) \mid \mathbf{x} \in \Phi\} \tag{20}$$

**Theorem III-C4.** The surface $V^{(L)}(\mathbf{x}) = v^{(L)}$ is completely contained in $\Omega$.

**Remark**: Note that even though there is no theoretical difficulty in estimate the stability boundary from (19)-(20), there could be a huge computational burden for systems with high dimensions [15]. Thus, the above analysis is applied to each of the mode-decoupled systems in (8) rather than the original $N$-dimensional system in (1). In theory, the Lyapunov function with infinite terms in (17) is independent of the choice of the function $\varphi$ used in the Zubov's equation (15). However, when a finite number of terms are kept in (18), the choice of the function $\varphi$ influences the rate of convergence. In addition, the intuition that keeping more terms will always give better accuracy is unnecessarily true even for an SMIB system [16]. The optimal selection of the function and the order $L$ is not a focus of this paper but deserves further investigations [17]. In the rest of this paper, we choose $\varphi$ and $L$ based on our experience on extensive case studies and use them for all test cases.

## IV. NONLINEAR MODAL DECOUPLING BASED POWER SYSTEM TRANSIENT STABILITY ANALYSIS

Although there are no theoretical limitations in applying the nonlinear modal decoupling to any sizes of multi-machine power systems, the computation burden will be a limiting factor for large-scale power systems. This section first presents the direct application of NMD based TSA for small-scale multi-machine power systems, and then introduces an idea on extending such analysis to large-scale power systems.

### A. Power system modeling

Consider an $m$-machine power system in the classical model, whose generators are represented by the second-order swing equation and loads by constant impedances:

$$\ddot{\delta}_i + \frac{\varsigma_i}{M_i}\dot{\delta}_i + \frac{\omega_s}{M_i}(P_{mi} - P_{ei}) = 0 \tag{21}$$

$$P_{ei} = E_i^2 g_i + \sum_{j=1, j\neq i}^{m}\left[a_{ij}\sin(\delta_i - \delta_j) + b_{ij}\cos(\delta_i - \delta_j)\right] \tag{22}$$

where $i \in \{1,2,\ldots,m\}$, $\delta_i$, $P_{mi}$, $P_{ei}$, $E_i$, $M_i$ and $\varsigma_i$ respectively represent the absolute rotor angle, mechanical power, electrical power, electromotive force, inertia constant and damping constant of machine $i$, and $g_i$, $a_{ij}$, and $b_{ij}$ represent network parameters including all loads.

Assume that the system has a uniform damping, i.e. constant $\varsigma_i/M_i$ for all $i$. It has been shown that (i) the oscillatory dynamics and the stability of the system are dominated by the relative motions among different machines [18][19]; and (ii) those relative motions can always be represented by an ($m$-1)-oscillator system [20].

Denote $\Delta = [\boldsymbol{\delta}^T \; \dot{\boldsymbol{\delta}}^T]^T$ as the state vector. Then, the first-order differential equaitons of the system in (21)-(22) have the form
$$\dot{\Delta} = \mathbf{f}_0(\Delta) \tag{23}$$

Apply a transformation matrix $R$, whose columns are right eigenvectors of $\mathbf{f}_0$'s Jacobian matrix, to both sides of (23) to obtain its modal space representation in (24), where $\mathbf{y} = [y_1 \; y_2 \; \ldots \; y_{2m}]^T$ is the new state vector.

$$\dot{\mathbf{y}} = R^{-1}\mathbf{f}_0(R\mathbf{y}) \triangleq \mathbf{g}(\mathbf{y}), \quad \text{where } \mathbf{y} = R^{-1}\Delta \tag{24}$$

Without loss of generality, let $y_1$, $y_2$, ..., $y_{2m-2}$ represent the relative motions of the system, and $y_{2m-1}$ and $y_{2m}$ represent the mean motions, which are non-oscillatory dynamics that all generators are moving together. It has been proved in [20] that the relative motions can be represented by an ($m$-1)-oscillator system consisting of different equations about $y_1$, $y_2$, ..., $y_{2m-2}$. Thus, the first ($2m$-2) equations of (24) are the model in (1) with $N = (2m-2)$, to which the NMD will be applied.

### B. NMD based TSA for small multi-machine power systems

For a power system with not many machines, the NMD can be applied directly to the entire system to generate $m$-1 independent second-order systems in the form of (8) with polynomial nonlinearities up to a desired order $k$, as shown by the following procedure (named "**NMD-TSA 1**"):

*Step 1:* Given an $m$-machine system represented by (23), derive the modal space representation (24).
*Step 2:* Assume uniform damping to obtain a unique ($m$-1)-oscillator system described by the first $2m$-2 equations in (24).
*Step 3:* Apply NMD to this ($m$-1)-oscillator system to obtain $m$-1 decoupled second-order systems in the form of (8) that respectively correspond to $m$-1 oscillatory modes.
*Step 4:* Apply a stability analysis in Section III to find the stability boundary of each of these ($m$-1)-second-order systems.
*Step 5:* For the original system, i.e. (24) or (21)-(22), given any trajectory subjected to a disturbance of interest, transform it to the decoupled coordinates using transformations (2) and (7), and visualize the transformed trajectory. If the trajectory does not exceed the stability boundary obtained in step 4 for any mode, the system subject to that disturbance is assessed to be stable; otherwise, it is considered unstable.

The stability boundary of each mode-decoupled system actually represents a portion of the stability boundary of the original system that is projected to the decoupled coordinates about one mode.

### C. NMD based TSA for large multi-machine power systems

For a large-scale multi-machine power system, the above **NMD-TSA 1** could be computationally expensive. It is often observed that when a power system is subject to a disturbance, usually only a few modes are significantly excited while the rest of the modes are either quiescent or less influential in stability of the system. Thus, a large-scale system can be reduced to a smaller system only about the dynamics associated with selected modes.

Re-write the first $2m$-2 equations in (24) as (25), which partitions $2m$-2 equations into one group for the modes of interest (denoted with subscript "int") and the other group for the rest of modes (denoted with subscript "non").

$$\begin{pmatrix} \dot{\mathbf{y}}_{\text{int}} \\ \dot{\mathbf{y}}_{\text{non}} \end{pmatrix} = \begin{pmatrix} \mathbf{g}_{\text{int}}(\mathbf{y}_{\text{int}}, \mathbf{y}_{\text{non}}) \\ \mathbf{g}_{\text{non}}(\mathbf{y}_{\text{int}}, \mathbf{y}_{\text{non}}) \end{pmatrix} \quad (25)$$

Ignore the dynamics with the second group, i.e. (26). Then, eq. (25) is reduced to (27) containing only the modes of interest.

$$\begin{cases} \dot{\mathbf{y}}_{\text{non}} = \mathbf{0} \\ \mathbf{y}_{\text{non}} = \mathbf{0} \end{cases} \quad (26)$$

$$\dot{\mathbf{y}}_{\text{int}} = \mathbf{g}_{\text{int}}(\mathbf{y}_{\text{int}}, \mathbf{0}) \quad (27)$$

Then, apply NMD to (27) and perform a stability analysis on each of the resulting mode-decoupled systems to obtain TSA results. The above enhanced TSA procedure for large-scale multi-machine systems is named "**NMD-TSA 2a**".

**Remark**: The **NMD-TSA 2a** procedure may introduce errors of TSA in two aspects. First, to reduce the large-scale system, dynamics with most of modes that do not impact stability are neglected, and hence the stability boundary estimated with respect to the modes of interest may not exactly match the true boundary. Second, inter-modal terms of orders > $k$ are not considered in the decoupled systems but they may become unneglectable when the system state is far from the SEP. Therefore, even if this procedure judges a post-disturbance trajectory to be stable, there is still a possibility that the original system has exits the true stability boundary due to influences from the ignored modes and high order inter-modal terms.

To address the above issue, define a shrinking ratio $r$ in (28) to reduce the stability boundary about the $i$th mode in (27) based on estimates of modal energies by (29)-(31). Here, we assume that the speed deviation of any generator rotor can be represented by a sum of sinusoids as shown in (31) about excited modes in (27). The representation (31) can be estimated by modal analysis tools, e.g. Prony and Matrix pencil methods, on stable trajectories.

$$r_i \triangleq E_i / E_{\text{all}} \quad (28)$$

$$E_i \triangleq a \sum_{j=1}^{m} H_j A_{ji}^2 \quad (29)$$

$$E_{\text{all}} \triangleq \sum_i E_i \quad (30)$$

$$\Delta \omega_j(t) = A_{ji} e^{\sigma_i t} \cos(\Omega_i t + \phi_{ji}) + \sum_{k \neq i} A_{jk} e^{\sigma_k t} \cos(\Omega_k t + \phi_{jk}) \quad (31)$$

The procedure that additionally applies the shrinking ratio to the stability boundary about each mode is named "**NMD-TSA 2b**" for comparison purposes. For instance, the estimated stability boundaries by the aforementioned first integral method and the Zubov's method become $V(w_{2i-1}, w_{2i}) = r_i V(0, w^*_{2i})$ and $V^{(L)}(\mathbf{x}) = r_i v^{(L)}$, respectively. Note that calculation of the shrinking ratio requires a stable trajectory and is contingency-dependent.

To summarize, NMD-based TSA for general power systems may contain three types of error: (i) truncation errors due to ignoring high-order power series terms of (1), (ii) model decoupling errors due to ignoring inter-modal terms of orders > $k$ of (2), and (iii) the estimation errors of the stability boundary. Ref. [11] investigated the first type of errors and suggests that a third order polynomial truncation be used for both conservative stability assessment and moderate computational burden, which will be adopted in case studies.

## V. CASE STUDIES

The first test is performed on an SMIB power system to investigate the third type of errors by comparing the stability boundaries estimated by the first integral method and Zubov's method to the reference result calculated by time simulations. The second test is on the 9-bus power system and investigates the second type of errors by comparing three procedures: **NMD-TSA 1**, **2a** and **2b**. The third test is conducted on a simplified NPCC 140-bus power system to show the potential of the NMD based TSA for a large power system. Finally, to test to what extent the NMD based TSA result, which is based on classical model, can work for the stability analysis with the associated detailed model, the detailed NPCC 140-bus power system is simulated with the same contingency in the third test, and visualize the marginal stable and unstable trajectories in the decoupled systems in the third test.

### A. Test on an SMIB system

Consider a general SMIB power system whose swing equation is shown in (32) [21].

$$\begin{cases} \dot{\delta} = \omega \\ \dot{\omega} = \dfrac{P_{\max} \omega_s}{2H}(\sin \delta_s - \sin(\delta + \delta_s)) - \dfrac{D}{2H}\omega \end{cases} \quad (32)$$

where $\delta$ is the relative rotor angle about its steady state value $\delta_s = 15°$, $\omega_s$ is the synchronous angular speed = $2\pi \times 60$ rad/s, $D=1$ and $H=3$s are the damping and inertia constants, respectively, and $P_{\max}=1.7$ p.u. is the maximum power transfer.

The third-order polynomial approximate of (32) at the origin (the SEP) is shown below.

$$\begin{cases} \dot{\delta} = \omega \\ \dot{\omega} = -0.1667\omega - 103.2\delta + 13.82\delta^2 + 17.2\delta^3 \end{cases} \quad (33)$$

Using the first integral method with the assumptions in (10) and (11), the system (33) is reduced to (34) and the resulting Lyapunov function is (35).

$$\begin{cases} \dot{\delta} = \omega \\ \dot{\omega} = -103.2\delta + 13.82\delta^2 + 17.2\delta^3 \end{cases} \quad (34)$$

$$V(\omega, \delta) = 0.5\omega^2 + 51.59\delta^2 - 4.608\delta^3 - 4.299\delta^4 \quad (35)$$

The two closest UEPs are $\delta_{\text{uep1}} = 2.0803$rad and $\delta_{\text{uep2}} = -2.8842$rad with energies $V(0,\delta_{\text{uep1}}) = 101.3$ and $V(0,\delta_{\text{uep2}}) = 242.2$. Thus, $V(0,\delta_{\text{uep1}})$ is defined as the critical energy and the corresponding stability boundary is estimated as (36).

$$0.5\omega^2 + 51.59\delta^2 - 4.608\delta^3 - 4.299\delta^4 = 101.3 \quad (36)$$

In the Zubov's method based approach, choose $\varphi(\omega,\delta)$ as

$$\varphi(\omega,\delta) = 0.0002\omega^2 + 0.001\delta^2 \quad (37)$$

Let $L=16$, which is adopted in all case studies with Zubov's method. Due to the size of the solution for $L=16$, we only illustrate the resulting Lyapunov function for $L=5$, i.e. $V^{(5)}$:

$$V^{(5)}(\omega,\delta) = 6.291\times10^{-4}\omega^2 + 9.692\times10^{-6}\omega\delta + 0.06491\delta^2$$
$$+8.386\times10^{-9}\omega^3 + 4.193\times10^{-9}\omega^2\delta + 1.299\times10^{-6}\omega\delta^2$$
$$-5.797\times10^{-3}\delta^3 - 1.771\times10^{-7}\omega^4 + 7.75\times10^{-9}\omega^3\delta \quad (38)$$
$$-3.654\times10^{-5}\omega^2\delta^2 + 1.16\times10^{-6}\omega\delta^3 - 7.294\times10^{-3}\delta^4$$
$$+3.172\times10^{-11}\omega^5 + 2.811\times10^{-11}\omega^4\delta + 8.192\times10^{-9}\omega^3\delta^2$$
$$+3.269\times10^{-6}\omega^2\delta^3 + 4.281\times10^{-7}\omega\delta^4 + 3.369\times10^{-4}\delta^5$$

All following calculations are based on $V^{(16)}$. The critical energy in (20) is found to be $v^{(16)} = 0.1142$ and the stability boundary is estimated as (39).

$$V^{(16)}(\omega,\delta) = v^{(16)} = 0.1142 \quad (39)$$

The stability boundaries of the system (33) determined by first integral, Zubov's method and time simulations are compared in Fig.1, which shows that the boundaries by first integral and Zubov's method are completely inside and fairly close to that by time simulations. Fig.1 also compares the stability boundaries of (33) and (32) by time simulations and shows that the third-order truncation in (33) consistently gives conservative stability analysis results for the system in (32).

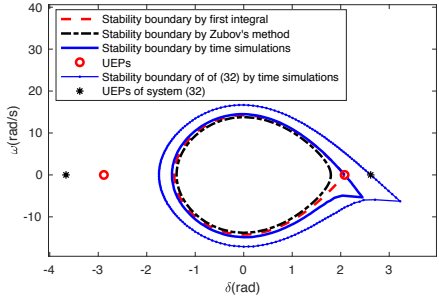

Fig.1. Stability boundaries estimated by different approaches.

### B. Test on a 3-machine 9-bus power system

Consider the 3-machine, 9-bus system [22] in the classical model [24], whose one-line diagram and the system equations corresponding to (23), (24), (5) and (8) after the studied contingency in this paper can be found as Fig.1, equations (56), (57), (59) and (61) in ref. [8]. Consider a temporary three-phase fault at bus 5 with the line 5-7 tripped upon the clearance of the fault, and the TSA is performed for the post-contingency system. The post-contingency system has two electromechanical modes, i.e. 0.96Hz and 2.05Hz, such that we have two mode-decoupled systems in the form of (8).

First, we apply **NMD-TSA 1** to analyze the two mode-decoupled systems. Fig.2(a) and Fig.2(b) show the identified stability boundaries by first integral, Zubov's method and time simulations (legends are the same as Fig.1), where a marginally stable trajectory with fault cleared after 8 cycles and a marginally unstable trajectory with fault cleared after 9 cycles are also drawn. Comparing Fig.2(a) and Fig.2(b) we can see that under this contingency, the 0.96Hz mode is dominant while the 2.05Hz mode is quite stable and only exhibits small dynamics even when the system becomes unstable. For each of the two modes, the BOA from Zubov's method is completely contained by that from time simulations, which means Zubov's method consistently provides conservative stability analysis results. However, a small portion of the BOA from the first integral method is outside of that from time simulations, which indicates that the first integral method may fail to identify some unstable cases. The time domain results of the two marginal cases are provided in Fig.3, whose projections in the two mode-decoupled coordinates in Fig.2(a) and Fig.2(b) can be clearly differentiated by the NMD based analysis, confirming the accuracy of the **NMD-TSA 1**.

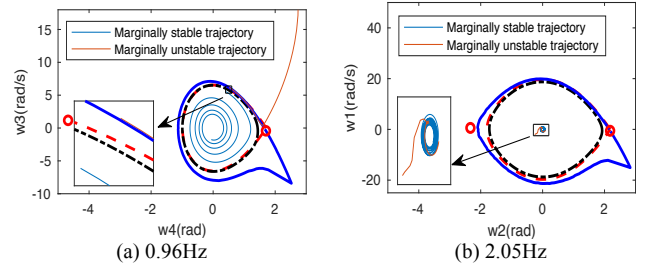

Fig.2. Stability boundaries of mode-decoupled systems by NMD-TSA 1.

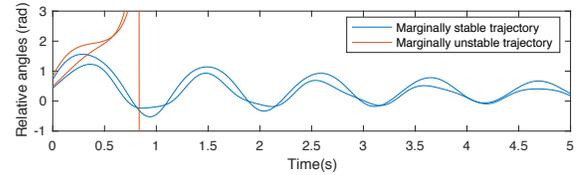

Fig.3. Marginally stable and unstable relative angles w.r.t. generator 1.

To validate **NMD-TSA 2a**, the simplified modal representation in the form of (27) is formulated for the 0.96Hz mode only, since it has dominant dynamics. The simplified formulation is then used to identified the stability boundary as shown in Fig.4(a), which is almost the same as Fig.2(a). This implies that ignoring the dynamics with the 2.05Hz mode does not introduce significant errors to stability analysis on the 0.96Hz mode. Such simplification is also valid for stability analysis on the 2.05Hz mode with the dominant 0.96Hz mode ignored as shown in Fig.4(b).

In addition, the two modes can be selected simultaneously to be analyzed by NMD-TSA 2a. In this case, **NMD-TSA 2a** is the same as **NMD-TSA 1**.

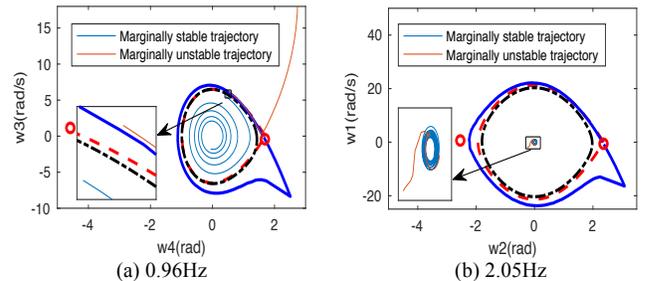

Fig.4. Stability boundaries of mode-decoupled systems by NMD-TSA 2a.

To check the performance of **NMD-TSA 2b**, we first apply NMD to both modes. Then, the marginally stable trajectory is adopted to calculate the shrinking ratio for each mode, which is then used to determine the shrunk BOAs by first integral and Zubov's method. The results are shown in Fig.5(a) and Fig.5(b) with $r_1=0.914$ and $r_2=0.086$, respectively. Comparing Fig.5(a) to Fig.2(a) or Fig.4(a), the estimated stability boundary for the dominant 0.96Hz mode is slightly changed. From Fig.5(b), the BOA for the 2.05Hz mode is significantly shrunk. As expected, more conservative stability analysis is resulted. Under the studied contingency, it is clear that the system is quite stable with the 2.05Hz mode and is more likely to lose stability with the 0.96Hz mode.

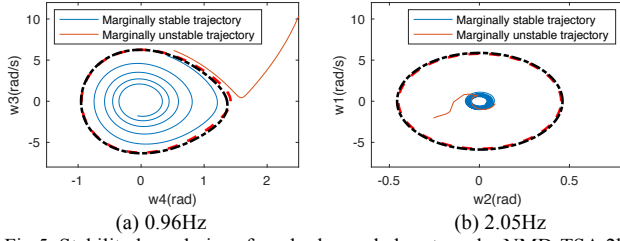

Fig.5. Stability boundaries of mode-decoupled systems by NMD-TSA 2b.

### C. Test on NPCC 140-bus power system with classical model

The third test is performed to show that NMD based TSA has potentials for large power systems. A simplified NPCC system is adopted for this purpose [25][26], which contains 48 generators and 140 buses. A temporary three-phase fault is added at bus 13 and cleared after a certain time without disconnecting any line. The critical clearing time of this contingency is 0.16 second and the resulting post-contingency response is shown in Fig.6. By calculating the modal energies according to the definition in (29)-(31), it is found that this fault only largely excites a few modes, as indicated by Table I. When the fault duration increases to 0.17s, the system will lose its stability as shown in Fig.7, where all rotor angles are divided into two clusters. The diagram of NPCC 140-bus system and the two clusters when losing stability is shown in Fig. 7.

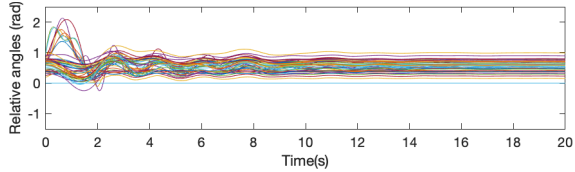

Fig.6. Marginally stable relative rotor angles w.r.t. generator 78.

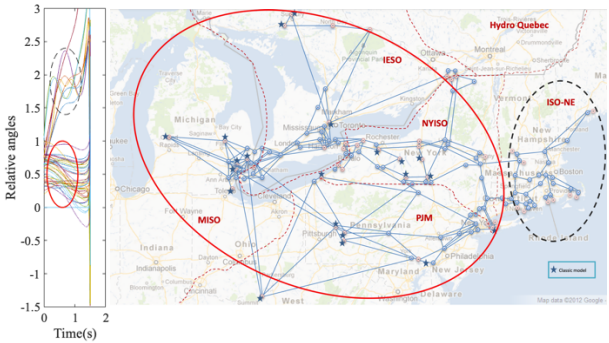

Fig.7. One-line diagram of the 140-bus NPCC power system (right) and the marginally unstable relative rotor angles w.r.t. generator 78 (left).

The **NMD-TSA 2b** is applied to the top-five largely excited modes. The stability boundaries estimated by **NMD-TSA 2b** with first integral or Zubov's method about five modes are shown in Fig.8, respectively. The legends are the same as those in Fig.1 and Fig.5. On one hand, it should be noted that the complex behaviors of generators shown in Fig.6 become much simpler to understand by Fig.8, which clearly show whether the system is going back to the SEP or not upon the fault clearance. On the other hand, the estimated stability boundary of each mode-decoupled system together with the system trajectory provides useful stability margin information.

For instance, the four modes in Fig.8(a)(b)(c)(e) are quite stable even though they are excited to exhibit noticeable dynamics; the 0.60Hz mode in Fig.8(d) is more likely to transit to instability modes because the post-contingency state of the system is close to the boundary. In addition, the two clusters also match the mode shape of the 0.60Hz mode.

TABLE I
MODAL ENERGY UNDER THE STUDIED CONTINGENCY

| $f_i$(Hz) | $E_i^*$ | $f_i$(Hz) | $E_i$ | $f_i$(Hz) | $E_i$ | $f_i$(Hz) | $E_i$ |
|---|---|---|---|---|---|---|---|
| 0.38 | 1 | 1.28 | $<10^{-3}$ | 1.57 | $<10^{-4}$ | 1.69 | $<10^{-6}$ |
| 0.26 | 0.51 | 1.56 | $<10^{-3}$ | 1.58 | $<10^{-5}$ | 1.99 | $<10^{-6}$ |
| 0.53 | 0.17 | 0.96 | $<10^{-3}$ | 1.40 | $<10^{-5}$ | 1.45 | $<10^{-6}$ |
| 0.60 | 0.12 | 1.04 | $<10^{-3}$ | 1.68 | $<10^{-5}$ | 2.51 | $<10^{-6}$ |
| 0.47 | 0.02 | 0.83 | $<10^{-3}$ | 1.28 | $<10^{-5}$ | 1.70 | $<10^{-7}$ |
| 2.44 | 0.01 | 0.95 | $<10^{-3}$ | 1.20 | $<10^{-5}$ | 1.41 | $<10^{-7}$ |
| 1.27 | $<10^{-2}$ | 0.91 | $<10^{-3}$ | 1.63 | $<10^{-5}$ | 1.51 | $<10^{-8}$ |
| 1.14 | $<10^{-2}$ | 1.55 | $<10^{-3}$ | 2.14 | $<10^{-5}$ | 1.87 | $<10^{-8}$ |
| 1.41 | $<10^{-2}$ | 1.38 | $<10^{-4}$ | 2.09 | $<10^{-5}$ | 1.85 | $<10^{-9}$ |
| 0.72 | $<10^{-2}$ | 1.78 | $<10^{-4}$ | 1.33 | $<10^{-6}$ | 1.69 | $<10^{-10}$ |
| 0.70 | $<10^{-3}$ | 1.72 | $<10^{-4}$ | 2.06 | $<10^{-6}$ | 1.35 | $<10^{-33}$ |
| 1.08 | $<10^{-3}$ | 1.17 | $<10^{-4}$ | 1.78 | $<10^{-6}$ | | |

* All the model energies are normalized by the energy of the 0.38Hz mode

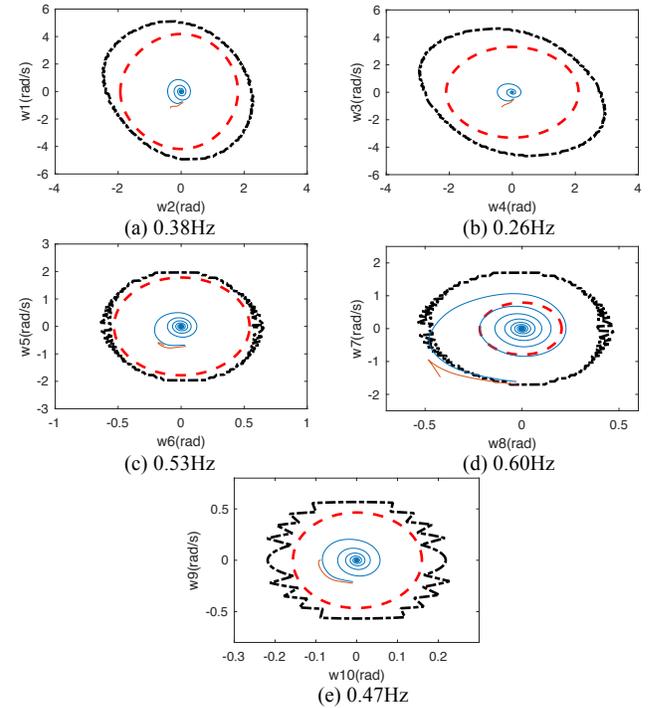

Fig.8. Shrunk stability boundaries of mode-decoupled system by NMD-TSA 2b. Trajectories are obtained using classical model.

Therefore, mode-by-mode stability information can be provided by NMD based TSA for a large power system and is valuable for identification of the most vulnerable grid interfaces for remedial control actions. However, by no means it can replace conventional simulation-based TSA tools since the results from NMD based TSA are still approximate, as illustrated in Fig.8(d), where the marginally stable trajectory has a small portion outside the BOA while the marginally unstable trajectory has a small portion inside.

### D. Test on NPCC 140-bus power system with detailed models

The purpose of the fourth test is to show to what extent the NMD based TSA result based on the classical model can work for the stability analysis with the associated detailed model.

The detailed NPCC 140-bus power system [26] is simulated with the same contingency in the third test, and visualize the marginal stable and unstable trajectories in the decoupled systems in the third test.

Fig.9 shows the projection of the marginal stable and unstable system trajectories to each mode. The legends are the same as those in Fig.1 and Fig.5. The stability analysis result is compatible with that of the third test. The four modes in Fig.9(a)(b)(c)(e) are quite stable even though they are excited to exhibit noticeable dynamics; the 0.60Hz mode in Fig.9(d) is more likely to transit to instability modes because the post-contingency state of the system is close to the boundary. Hence, the NMD based TSA has the potential to work for the detailed model situation.

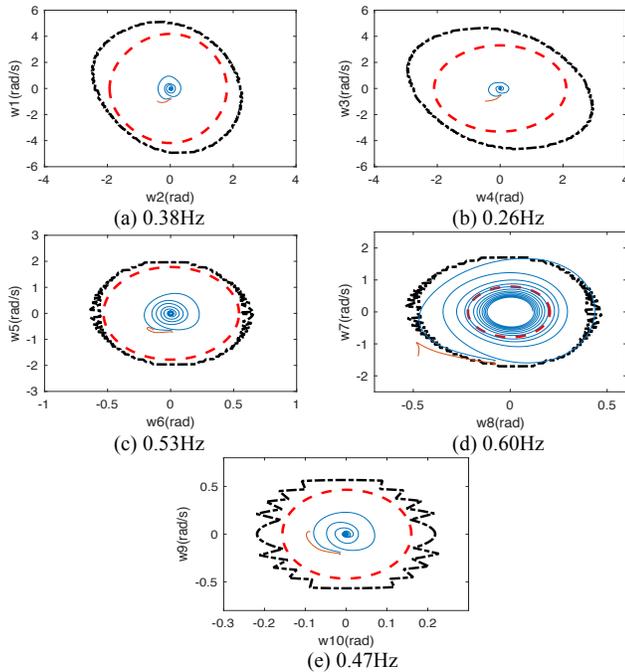

Fig.9. Shrunk stability boundaries of mode-decoupled system by NMD-TSA 2b. Trajectories are obtained using detailed model.

## VI. CONCLUSIONS AND FUTURE WORK

For general multi-oscillator systems including multi-machine power systems in the classical model, the previously proposed nonlinear modal decoupling (NMD) approach provides an approximate representation for stability analysis, i.e. a number of decoupled nonlinear one-degree-of-freedom oscillators. This paper adopts the NMD approach and analyzes the transient stability of each of the decoupled nonlinear oscillators to infer the transient stability of the original multi-machine power systems by using the Lyapunov function theory. The proposed idea is validated on a single-machine-infinite-bus power system, the 9-bus power system, and the simplified and detailed NPCC 140-bus power system. Test results show that the NMD based analysis has a potential to assess the transient stability and visualize the modal dynamics of multi-machine power systems.

Future works will consider more detailed power system models. Online application of NMD based TSA will also be studied, which has the potential to generate the stability boundary every 15min. Another alternative is to offline produce for a variety of operating conditions and then online select the best fit for monitoring purpose.